\documentclass[aps,pra,twocolumn,showpacs,superscriptaddress,nofootinbib
]{revtex4-1}

\usepackage{ae,aecompl} % for smooth PDF font

\usepackage{amsmath}  
\usepackage{amsthm}  
\usepackage{amssymb}  
\usepackage{color}

\usepackage{longtable}
\usepackage{multirow}
\usepackage{mathrsfs}
\usepackage{braket}
\usepackage{xfrac}

\usepackage{hyperref}
\hypersetup{
	colorlinks=true,
	linkcolor=blue,     
	citecolor=green,    
	urlcolor=magenta       
}

\let\savedegree\corresponds
\let\corresponds\relax
\usepackage{mathabx}
\usepackage[bb=boondox]{mathalfa}
\let\corresponds\savedegree

\usepackage{float}
\usepackage{bm}
\usepackage[percent]{overpic}
\usepackage{mathrsfs}
\usepackage[table,dvipsnames]{xcolor}
\usepackage[ruled]{algorithm2e}

\usepackage[caption=false]{subfig}
\usepackage{graphicx}

\definecolor{green}{rgb}{0.2, 0.7, 0.2}
\definecolor{blue1}{rgb}{0.3,0.3,1}

\newcommand{\ii}{\texttt{i}} %for iota = \sqrt{-1}

\begin{document}
\title{Interferometric Neural Networks}

\author{Arun Sehrawat}
\email[]{arunsehrawat2@gmail.com}
\affiliation{QpiAI India Pvt. Ltd., Hub 1 SEZ Tower, Karle Town Centre, Nagavara, Bangalore 560045, India}

\date{October 25, 2023}

%===========================================

\begin{abstract}
	
On the one hand, artificial neural networks have many successful applications in the field of machine learning and optimization. On the other hand, interferometers are integral parts of any field that deals with waves such as optics, astronomy, and quantum physics. Here, we introduce neural networks composed of interferometers and then build generative adversarial networks from them.
Our networks do not have any classical layer and can be realized on quantum computers or photonic chips.
We demonstrate their applicability for combinatorial optimization, image classification, and image generation. 
For combinatorial optimization, our network consistently converges to the global optimum or remains within a narrow range of it.  
In multi-class image classification tasks, our networks achieve accuracies of ${93\%}$ and ${83\%}$. 
Lastly, we show their capability to generate images of digits from 0 to 9 as well as human faces.

\end{abstract}

%===========================================

\maketitle

%===========================================
\section{Introduction}\label{sec:Intro}

Artificial neural networks (NNs) \cite{LeCun15,Goodfellow16, Aggarwal23} 
have demonstrated remarkable success in a wide range of applications in various domains such as autonomous vehicles, healthcare, finance, robotics, gaming, over-the-top media services, and chatbots. Some of the applications include image classification \cite{Lecun89,Lecun98,Krizhevsky12}, natural language processing, speech recognition, robotic control systems, recommendation systems, and image generation \cite{Goodfellow14,Radford15,Arjovsky17,Gulrajani17}.
Convolutional NNs (CNNs) were introduced in \cite{Lecun89,Lecun98} and further improved in seminal works like \cite{Krizhevsky12} for image classification. Similarly, generative adversarial networks (GANs) made their debut in \cite{Goodfellow14} and underwent notable advancements in subsequent works, including \cite{Radford15,Arjovsky17,Gulrajani17} for image generation.
In this paper, we introduce interferometric NNs (INNs) consisting of interferometers in Sec.~\ref{sec:INN} and then interferometric GANs (IGANs) composed of INNs in Sec.~\ref{sec:IGAN}.

Interferometers play a pivotal role in metrology, astronomy, optics, and quantum physics. Feynman, in his renowned lectures, introduced quantum (wave-particle) behavior through double-slit (two-path) interference experiments \cite{Feynman64}.
The wave-particle duality, a subject of the famous Bohr-Einstein debates, has been further investigated by numerous scientists using two- and multi-path interferometers (for instance, see \cite{Wootters79,Greenberger88,Englert96,Durr98a, Durr98b, Englert08, Coles16, Qureshi17}). Our INNs are made of sequences of such
multi-path interferometers, where each interferometer is a combination of 
beamsplitters (or fiber couplers) and a phase shifter \cite{Englert96, Englert08, Coles16, Weihs16}.
A beamsplitter is represented by a discrete Fourier transformation, 
and the phase shifters hold all the learnable parameters of an INN.
In principle, an INN can work with classical or quantum waves.
However, a sequence of interferometers can be viewed as a parameterized quantum circuit (PQC), where a Fourier transformation can be implemented exponentially faster than its classical counterpart thanks to Peter Shor \cite{Shor96}.

The PQCs \cite{Benedetti19} are integral components of the variational quantum algorithms (VQAs), including the variational quantum eigensolver (VQE) \cite{Peruzzo14,McClean16} and its variants \cite{Malley16,Motta18,Lee19,Matsuzawa20} for determining molecular energies and the quantum approximate optimization algorithm (QAOA) \cite{Farhi14} for solving combinatorial optimization problems by recasting them as energy minimization in Ising spin glass systems \cite{Lucas14}.
Recognizing that fully fault-tolerant quantum computers may still be decades away, VQAs are harnessing the capabilities of today's noisy intermediate-scale quantum (NISQ) computers \cite{Preskill18} by providing a framework to 
build a wide range of applications.
Some of these applications include finding ground \cite{Peruzzo14,McClean16} and excited \cite{Higgott19, Nakanishi19}
energy states, combinatorial optimization \cite{Farhi14,Lucas14},
simulating quantum dynamics \cite{Li17,Yuan19,Cirstoiu20,Gibbs21,Kivlichan18},
solving a system of linear \cite{Bravo-Prieto19,Huang19,Xu21} and non-linear 
\cite{Lubasch20,Kyriienko21} equations, data classification \cite{Farhi18,Havlicek19,Franken20,Mitarai18,Huggins19,Grant18,Schuld19,Schuld20}, and generative tasks 
\cite{Romero17,
	Lloyd18,Demers18, Zoufal19,Romero19,Niu21,
	Huang21, Tsang22, Zhou23, Stein21, Chu22,
	Liu18, Benedetti19b, Coyle20, Rudolph22,
	Tian22}.
For an in-depth review of VQAs and NISQ algorithms, we refer readers to \cite{Cerezo21, Bharti22}.
An INN also belongs to the VQA family, as it can provide an ansatz for a VQA.

We employ INNs for combinatorial optimization and image classification in Sec.~\ref{sec:INN} and IGANs for image generation in Sec.~\ref{sec:IGAN}. 
The notion of quantum GANs was initially introduced in \cite{Lloyd18} and
numerically implemented in \cite{Demers18} using PQCs.
Subsequently, various quantum GANs have been proposed to generate samples from random distributions \cite{Zoufal19,Romero19,Niu21}, as well as to generate images \cite{Huang21, Tsang22, Zhou23, Stein21, Chu22}. Leveraging the Born rule, which provides a probability distribution from a (parameterized) quantum state, PQCs as Born machines \cite{Liu18, Benedetti19b,Coyle20, Rudolph22} are utilized for these generative tasks.

We provide a summary of our contributions at the beginning of the next two sections and an overarching conclusion and outlook in Sec.~\ref{sec:conclusion}.
One can find a dedicated Jupyter Notebook for each of the problems discussed in this paper within our GitHub repository \cite{Arun_GitHub}. These problems are tackled using INNs, which are classically simulated in the PyTorch machine-learning framework \cite{Paszke19}. PyTorch automatically manages the gradient of a loss or an energy function. For the training of each INN, we employ the Adam optimizer \cite{Kingma14}.

%===========================================
\section{INN}\label{sec:INN}

\begin{figure}
	\includegraphics[width=0.48\textwidth]{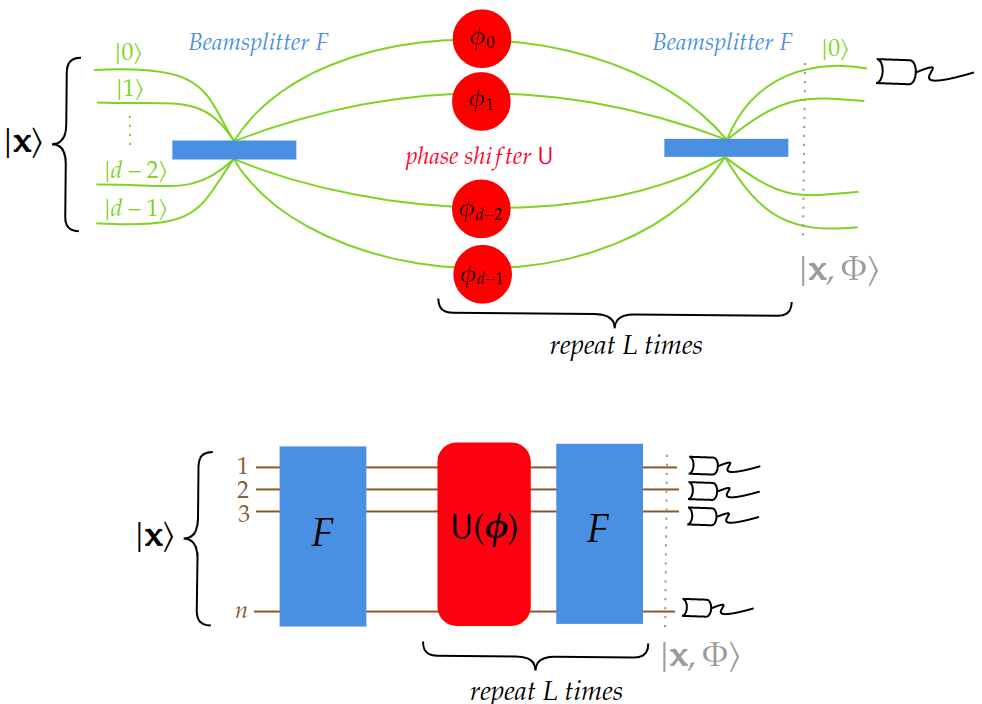}
	\caption{The top and bottom pictures show an INN and its equivalent PQC in the case of $d=2^n$, respectively.
	The INN is a sequence of $L$ interferometers, where the output from one is fed to the next. Each interferometer has $d$ distinct paths colored in green, beamsplitters (or fiber couplers) exhibited by blue-colored slabs, and a phase shifter shown by red disks in the top picture.
	In the bottom picture, brown wires depict $n$ qubits, and the
	beamsplitter and phase shifter are displayed by blue and red-rounded rectangles, respectively.
	An input to the INN or to the PQC is a ket ${|\textbf{x}\rangle}$ from \eqref{x-kets}, the beamsplitter $F$ and phase shifters $U(\boldsymbol{\phi}^l)$ are given in \eqref{FU}, and the output ${|\textbf{x}, \Phi \rangle}$ is read by the detectors shown in black.
	The whole setup is summarized in \eqref{p}.
}
	\label{fig:Int}
\end{figure}

\begin{figure}
	\includegraphics[width=0.45\textwidth]{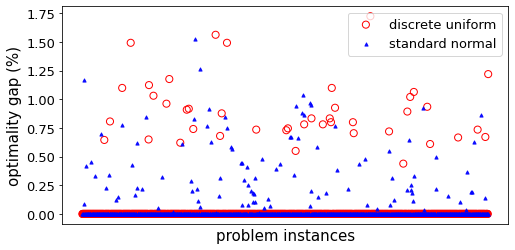}
	\caption{The plot illustrates optimality gaps for two distinct sets of QUBO problem instances, each containing one thousand instances for $n=17$ qubits. The red circles $({\color{red}\circ})$ and blue triangles $({\color{blue}\blacktriangle})$ represent the gaps for the two sets where the $\mathsf{Q}$ matrix is sampled from the discrete uniform and standard normal distributions, respectively. The INN consistently yields the global minimum, achieving a success rate of $95\%$ and $89\%$ for the uniform and normal distributions, respectively. In both cases, the optimality gaps remain under $2\%$.}
	\label{fig:opt_gap}
\end{figure}

\begin{figure}
	\includegraphics[width=0.4\textwidth]{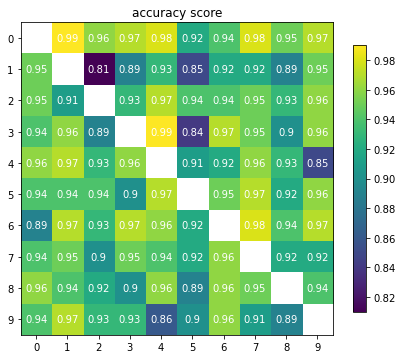}
	\includegraphics[width=0.4\textwidth]{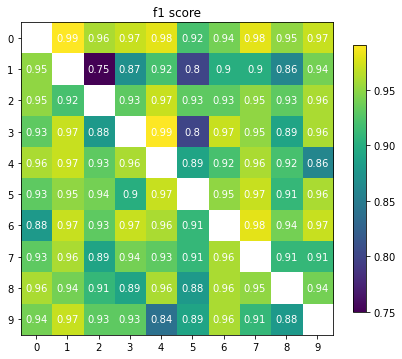}
	\caption{The top and bottom matrices display the accuracy and $\text{f}_1$ scores of trained INNs for binary classification problems. The entry in $a$th row and $b$th column corresponds to the image classification task of digit $a$ (negative class) against digit $b$ (positive class). It can be observed that the INN achieves accuracy ranging from $0.81$ to $0.99$ and $\text{f}_1$ scores between $0.75$ and $0.99$.}
	\label{fig:acc_f1}
\end{figure}

\begin{figure}
	\includegraphics[width=0.48\textwidth]{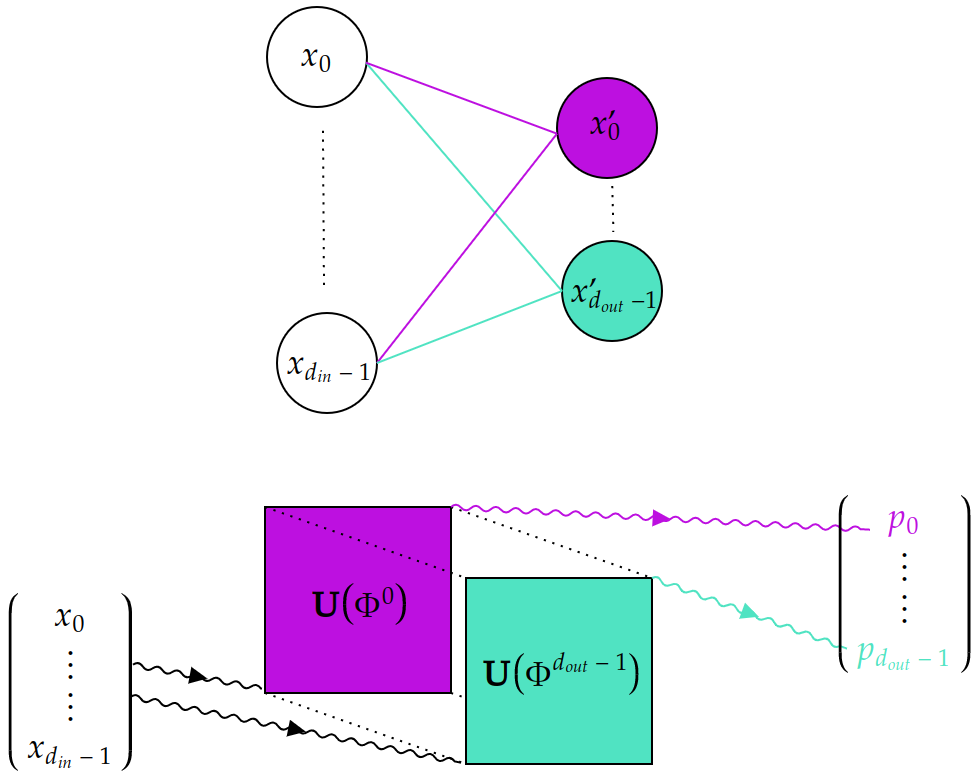}\vspace{2mm}
	\caption{The top and bottom pictures exhibit a linear layer [for example, given in \eqref{W, FU}] of a classical NN and an interferometric block [specified by \eqref{pt}] of an INN, respectively. By the color coding
		one can see a similar \emph{parallel} structure in both cases.
		The linear layer and INN block turn the input $\textbf{x}\in\mathbb{R}^{d_\text{in}}$ into the outputs 
		$\textbf{x}'\in\mathbb{R}^{d_\text{out}}$ and $\textbf{p}\in[0,1]^{\times d_\text{out}}$, respectively.
		In the bottom picture, the squares represent $d_\text{in}\times d_\text{in}$ \emph{independent} unitary matrices $\textbf{U}(\Phi^t)$, each of which is a sequence of interferometers as depicted in Fig.~\ref{fig:Int}. 
	}
	\label{fig:Int_blk}
\end{figure}

\begin{figure}
	\includegraphics[width=0.3\textwidth]{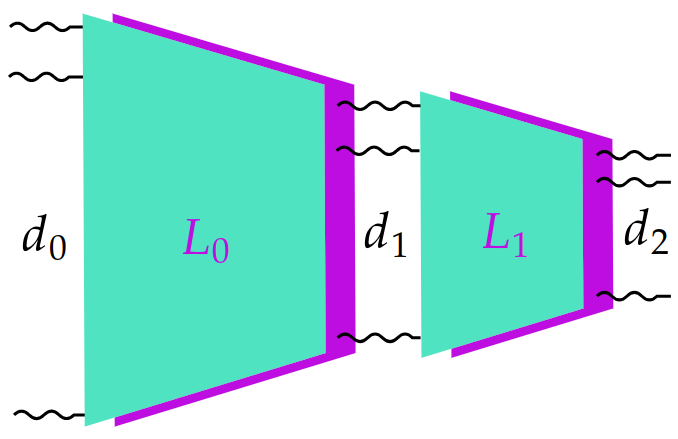}\vspace{2mm}
	\caption{
		The diagram illustrates an INN composed of two interferometric blocks, depicted as trapezoids. Each block shares a similar structure as shown in Fig.~\ref{fig:Int_blk}. From left to right, the INN transforms the dimensions as $d_0\rightarrow d_1\rightarrow d_2.$ The number of layers in the first and second blocks are denoted as $L_0$ and $L_1$, respectively.
		The architecture of this INN is characterized by
		$\textbf{d}=(d_0, d_1, d_2)$ and $\textbf{L}=(L_0, L_1)$.
}
	\label{fig:2Blks}
\end{figure}

In this section, our primary contributions are in 
\eqref{FU}--\eqref{dp}, \eqref{pt}, Figs.~\ref{fig:Int}--\ref{fig:2Blks}, Table~\ref{tab:acc_f1}, and \cite{Arun_GitHub}.
Here we introduce INNs through \eqref{FU}--\eqref{dp}, \eqref{pt},
Figs.~\ref{fig:Int}, \ref{fig:Int_blk}, and \ref{fig:2Blks}.
Initially applied to specific combinatorial optimization problems, we illustrate their performance in Fig.~\ref{fig:opt_gap}. Subsequently, we use them for image classification tasks, and their performance is detailed in Fig.~\ref{fig:acc_f1} and Table \ref{tab:acc_f1}.

The INN in Fig.~\ref{fig:Int} is made of a sequence of interferometers, where each interferometer has $d$ distinct paths represented by 
the computation (orthonormal) basis 
\begin{equation}
	\label{basis}
	\mathcal{B}=\big\{ |0 \rangle, \cdots, |d-1 \rangle\big\}
\end{equation}
of the associated Hilbert space.
Basically, this whole setup is a $d$-level quantum computer.
In the case of $d=2^n$, one can construct an equivalent PQC of $n$ qubits ($2$-level quantum systems) as shown in 
Fig.~\ref{fig:Int}.

In a machine-learning task, one has to feed the data to a NN, and the data is a collection of feature vectors $\textbf{x}=(x_0,\cdots,x_{d-1})\in\mathbb{R}^d$. Whereas, in the case of INN, we have to encode every \textbf{x} into an input state vector 
$|\textbf{x}\rangle$ through a unitary transformation. Such a process is called feature encoding \cite{Schuld20}.
Throughout the paper, we consider the \emph{amplitude} encoding  
\begin{equation}
	\label{x-kets}
	|\textbf{x}\rangle  = 
	\frac{1}{\|\textbf{x}\|}\sum_{j=0}^{d-1}  x_j\,|j \rangle\,,
\end{equation}
where $\|\textbf{x}\|$ is the Euclidean norm of $\textbf{x}$.

As shown in Fig.~\ref{fig:Int} following \cite{Englert96, Englert08, Coles16, Weihs16}, an interferometer has
two essential elements: \emph{beamsplitters} (or fiber couplers) and a \emph{phase shifter},
which, in our case, are mathematically described by the unitary operators 
\begin{align}
	\label{FU}
	F &= \frac{1}{\sqrt{d}}\sum_{k=0}^{d-1} \sum_{j=0}^{d-1} \omega^{\,kj}\,
	|k \rangle\langle j| 
	\quad\mbox{and} \nonumber\\
	U(\boldsymbol{\phi}) &= \sum_{j=0}^{d-1}  \text{e}^{\texttt{i}\phi_j}
	|j \rangle\langle j|\,,
\end{align}
respectively.
In \eqref{FU}, ${\omega=\text{e}^{\texttt{i}\frac{2\pi}{d}}}$ is a $d$th root of unity, $\texttt{i}=\sqrt{-1}$, and $\boldsymbol{\phi}=(\phi_0,\cdots,\phi_{d-1})\in\mathbb{R}^d$ carry phases.
Here the discrete Fourier transformation $F$ acts as a generalization of a symmetric $50/50$ beamsplitter for a $2$-path interferometer. It turns a single path $|j\rangle$ into 
an equal superposition of all the paths of $\mathcal{B}$.
The phase shifter $U(\boldsymbol{\phi})$ shifts the phase by $\phi_j$ in the $j$th path.
By the way, using the operators from \eqref{FU}, one can realize a phase encoding $|\textbf{x}\rangle_\text{ph} = U(\textbf{x})F|0\rangle$ of the data, which might be useful in some applications.
We want to emphasize that two layers of the Hadamard and diagonal unitary operations like $U$ are used in \cite{Havlicek19} to present a feature map that is hard to simulate classically. Additionally, parameterized diagonal unitary operations such as $U(\boldsymbol{\phi})$ have been employed to make a quantum simulation faster in \cite{Cirstoiu20,Gibbs21} (see also \cite{Kivlichan18}).

For every problem, we have executed $F$ in \cite{Arun_GitHub}, utilizing the fast Fourier transform available in PyTorch. However, its implementation can reach exponential speedup on a quantum computer due to Shor's algorithm for quantum Fourier transforms \cite{Shor96}. The phase shifter can be decomposed as a product of controlled-phase gates \cite{Kivlichan18}:  
\begin{align}
	\label{CP}
	U(\boldsymbol{\phi}) &= \prod_{j=0}^{d-1}\textsc{cp}(\phi_j)\,,
	\quad\mbox{where} \nonumber\\
	\textsc{cp}(\phi_j) &= 	\text{e}^{\texttt{i}\phi_j}|j \rangle\langle j| + 
	\big(I - |j \rangle\langle j|\,\big)
	= \text{e}^{\texttt{i}\phi_j|j \rangle\langle j|}
\end{align}
and $I$ is the identity operator.
In the case of $d=2^n$, with the binary representation of integers $j=\sum_{\mu=1}^n 2^{n-\mu}j_\mu$ and association $|j\rangle \equiv \otimes_{\mu=1}^n |j_\mu\rangle$, one can show that $\textsc{cp}$ is indeed a controlled-phase gate with $n-1$ control qubits. Although there are exponentially many $\textsc{cp}$ gates relative to $n$ in the decomposition, they can all, in principle, be simultaneously realized due to their commutativity. By the way, for $\phi_j=\pi$, $\textsc{cp}(\phi_j)$ becomes a quantum oracle of Grover's algorithm \cite{Grover97}.

Now, we have all the elements to describe the INN of Fig.~\ref{fig:Int}.
The INN is made of $L$ layers of interferometers, where the output from one is
fed to the next. An input to the first interferometer is $|\textbf{x}\rangle$ from \eqref{x-kets}, the output from the last one is
\begin{align}
	\label{p}
	|\textbf{x}, \Phi \rangle & = 
	\left(\prod_{l=1}^{L} F U(\boldsymbol{\phi}^{l}) \right) F\, |\textbf{x}\rangle\,,
	\quad\mbox{and} \nonumber\\
	p	&= |\langle 0|\textbf{x}, \Phi \rangle|^2
\end{align}
is the probability of getting $0$th outcome (that is, detecting quantum particles in the $0$th path). Equivalently, in the case of $n$ qubits, we have to perform measurements on all the qubits to get $p$ as $|0\rangle\equiv|0\cdots0\rangle$ in the binary representation.
Equations in \eqref{p} completely specify the INN in Fig.~\ref{fig:Int}, and all its learnable parameters are denoted by 
$\Phi=(\boldsymbol{\phi}^1,\cdots,\boldsymbol{\phi}^L)\in\mathbb{R}^{L\times d}$, which carries $L$ phase-vectors, one for each interferometer.

One can view the INN as a differentiable function 
$p: \mathbb{R}^{L\times d}\rightarrow [0,1]$
on the parameter space. 
In optimization or machine-learning tasks, PyTorch automatically manages the gradient computation of an energy or a loss function [given in \eqref{eng}, \eqref{loss}, and \eqref{loss_GAN}]
of $p$ with respect to parameter updates. Without such a framework, we would need to compute derivatives such as
\begin{align}
	\label{dp}
&\frac{\partial p}{\partial \phi^l_j} = 
\langle\textbf{x}|\cdots \partial_j U^\dagger(\boldsymbol{\phi}^l)
\cdots|0\rangle\langle0|\cdots U(\boldsymbol{\phi}^l)\cdots|\textbf{x}\rangle\ +\quad
\nonumber\\
&\qquad\quad\, \langle\textbf{x}|\cdots U^\dagger(\boldsymbol{\phi}^l)
\cdots|0\rangle\langle0|\cdots \partial_j U(\boldsymbol{\phi}^l)\cdots|\textbf{x}\rangle\,,
\ \mbox{where}
\nonumber\\
&\partial_j U(\boldsymbol{\phi}^l)  = \frac{\partial U}{\partial \phi^l_j} 
= \texttt{i}\,\text{e}^{\texttt{i}\phi^l_j } |j\rangle\langle j|
=\frac{\partial \textsc{cp}(\phi^l_j)}{\partial \phi^l_j} \,.
\end{align}
The second equation in \eqref{dp} expresses the derivative of the $l$th phase shifter 
with respect to the phase $\phi^l_j$, and it also turns out to be the derivative of the \textsc{cp} operator.  Similarly, one can get 
the derivative $\partial_j U^\dagger(\boldsymbol{\phi}^l)$ of the adjoint (denoted by $\dagger$) of $U$.

Now, we demonstrate applications of the INN first for quadratic unconstrained binary optimization (QUBO) and then for classification.
A $n$-variable QUBO problem is equivalent to finding a minimum energy eigenstate of an
associated $n$-qubit Hamiltonian (for many such problems, see \cite{Lucas14})
\begin{equation}
	\label{H}
	H = \sum_{\mu,\nu=1}^n \mathsf{Q}_{\mu\nu} P_\mu P_\nu\,,
\end{equation}
where $\mathsf{Q}$ is a $n\times n$ real matrix that defines the problem, and
the projector $P_\mu$ acts on the $\mu$th qubit as 
$P_\mu|j_\mu \rangle = j_\mu|j_\mu \rangle\,, j_\mu\in\{0,1\}$.
By the binary representation $j=\sum_\mu 2^{n-\mu}j_\mu$
and $|j\rangle \equiv \otimes_\mu |j_\mu\rangle$,
$H$ is diagonal in the computational basis $\mathcal{B}$ of \eqref{basis}.
The global minimum will be
$E_j = \langle j |H| j\rangle$ for some $j\in\{0,\cdots,d-1\}$.
Since $\{E_j\}$ does not have any structure like convex functions, finding the minimum out of exponentially ($d=2^n$) many possibilities is an NP-hard problem.

Several successful VQAs, including the VQE \cite{Peruzzo14,McClean16} and QAOA \cite{Farhi14}, have been proposed to find good approximate solutions. Their basic structure involves preparing a parametric state (referred as ansatz), computing and minimizing the energy expectation value, and obtaining an approximate solution from the optimized ansatz. 
A QAOA ansatz is constructed by sequentially applying the problem and a mixer unitary operators, whereas the VQE ansatz is generated by applying unitary coupled cluster operators to an initialized state (for more details, see \cite{McClean16,Cerezo21,Bharti22}). 
Our approach is similar as narrated next, with the key distinction being that our ansatz is derived from the INN of Fig.~\ref{fig:Int}.

We start with the input ket $|\textbf{x}\rangle=|0\rangle$ and reach 
the output ket $|0,\Phi\rangle\equiv|\Phi\rangle$ as per \eqref{p}.
Then we perform measurements to compute the energy expectation value $\mathcal{E}(\Phi)$ to minimize it over the parameter space and obtain
\begin{align}
	\label{eng}
\Phi_\text{sol} &= \underset{\Phi}{\text{argmin}}\ \mathcal{E}(\Phi)\,,
 \\
	\mathcal{E}(\Phi) &= \langle\Phi|H|\Phi \rangle 
	= \sum_{j=0}^{d-1} E_j|\langle j|\Phi \rangle|^2 = 
	\sum_{\mu,\nu=1}^n \mathsf{Q}_{\mu\nu} \langle P_\mu P_\nu\rangle\,\,.\nonumber
\end{align}
Usually, quantum and classical computers are put in a loop to compute and minimize
$\mathcal{E}$, respectively. To compute $\mathcal{E}$, one needs only $n^2$ expectation values $\langle P_\mu P_\nu\rangle$ of local observables.
After \eqref{eng}, we perform measurements on the optimized ansatz $|\Phi_\text{sol}\rangle$ in the computational basis $\mathcal{B}$ to get the most probable outcome
\begin{equation}
	\label{jsol}
	j_\text{sol} = \underset{j}{\text{argmax}}\ 
	|\langle j|\Phi_\text{sol}\rangle|^2
\end{equation}
as our solution. 
If the global minimum $E_\text{min} = \min_j \{E_j\}$ is known then one can compute
\begin{equation}
	\label{gap}
	\text{optimality gap} = \left|\frac{E_{j_\text{sol}} - E_\text{min}}{E_\text{min}}\right| \times 100\,,
\end{equation}
where $E_{j_\text{sol}}=\langle j_\text{sol} |H| j_\text{sol}\rangle$ corresponds to the obtained solution via the INN.

For $n=17$ qubits, to generate QUBO instances, we sampled $\mathsf{Q}$ from one of the two distributions: the uniform distribution over the set $\{-10,\cdots,10\}$ of integers and the standard normal distribution. From each distribution, we generated one thousand QUBO instances.
For each instance, we obtained $E_\text{min}$ through an exhaustive search over $d=2^{17}=131072$ possibilities and reported the optimality gap in Fig.~\ref{fig:opt_gap}. 
In the case of discrete uniform distribution, we achieved the global minimum in ${958}$ out of ${1000}$ instances, meaning that the gap was $\leq 10^{-10}$. 
Over all the instances, the mean and maximum gaps were $0.04\%$ and $1.72 \%$, respectively.
When using the standard normal distribution, we reached the global minimum in ${894}$ instances, and the mean and maximum gaps were $0.04\%$ and $1.52\%,$ respectively.
For each instance, we maintained a consistent configuration with a number of layers, $L=2$, and carried out $201$ epochs of training with a learning rate of $0.05$ and $\text{betas}=(0.5, 0.9)$ for the Adam optimizer. Further details can be found in \cite{Arun_GitHub}.

It will be interesting to see how the INN will perform if we increase the system size $n$. For this, we need \emph{quantum} hardware that can store ${|\Phi \rangle}$ and return $\mathcal{E}(\Phi)$. Beyond certain $n$, classical hardware cannot store exponentially many numbers $\langle j|\Phi \rangle$ for $j=0,\cdots,2^n-1$.
It is a true power for a quantum computer \cite{Peruzzo14}.
We are not exploring this direction any further but moving to the next problem.

In a $C$-class classification problem, our goal is to predict the true label  $y=(y_1,\cdots,y_C)\in\{0,1\}^C$ for a data point based on its features provided in $\textbf{x}$. 
When the true class is $c$, $y_c = 1$, and the remaining components of $y$ are set to zero.
During the training process of a NN for the problem, the objective is to minimize a specific loss function, such as the cross-entropy (negative log-likelihood)
\begin{equation}
	\label{loss}
	\mathcal{L} = -\frac{1}{N}\sum_{i=1}^N 
	\sum_{c=1}^C y^{[i]}_c \ln(p^{[i]}_c)
\end{equation}
within the parameter space.
The outer summation in \eqref{loss} is performed over a mini-batch $\{\textbf{x}^{[i]}, y^{[i]}\}_{i=1}^N$ containing $N$ data points.

Suppose the dataset contains only $C=2$ classes,
then one can interpret $p$ of \eqref{p} as the probability of a data point belonging to the positive class ($c=1$) and ${1-p}$ as the probability of it belonging to the negative class ($c=0$).
Consequently, the INN exhibited in Fig.~\ref{fig:Int} can be employed for binary classification.
To illustrate this, we have utilized the MNIST dataset \cite{MNIST}, which contains images of 0 to 9 handwritten digits. To create a 2-class classification problem, we gathered all the images of only two specific digits, denoted as $a$ and $b$, which are respectively labeled as the negative class $y=(1,0)$ and the positive class $y=(0,1)$. After the training of the INN, we evaluate its performance by using 
\begin{align}
	\label{acc_f1}
	\text{accuracy} &= \frac{\text{TP}+\text{TN}}%
	{\text{TP}+\text{TN}+\text{FP}+\text{FN}} 
	\quad\text{and}\nonumber\\
	\text{f}_1 &= \frac{2\,\text{TP}}{2\,\text{TP}+ \text{FP}+\text{FN}}
\end{align}
on the test set.
The results are presented in Fig.~\ref{fig:acc_f1} for every $a,b\in\{0,\cdots,9\}$ provided $a\neq b$.
Here, {TP}, {TN}, {FP}, and {FN} represent the counts of true positives, true negatives, false positives, and false negatives, respectively.
Note that if we were to interchange the labels of $a$ and $b$, then $\mathcal{L}$, the trained INN, accuracy, and $\text{f}_1$ score will be different. 
As a result, the matrices in Fig.~\ref{fig:acc_f1} are not symmetric around their diagonals. Various binary classifications have been carried out on the MNIST dataset in \cite{Farhi18, Franken20, Huggins19, Grant18, Schuld20} employing diverse PQCs. 
In our case, we achieve comparable performance, as illustrated in Fig.~\ref{fig:acc_f1}.

An image consists of $d=\mathsf{C}\times\mathsf{H}\times\mathsf{W}$ pixel values, where $\mathsf{C}$, $\mathsf{H}$, and $\mathsf{W}$ denote the number of channels, height, and width in pixels, respectively. Since the MNIST dataset contains black and white images, $\mathsf{C}=1$. 
To get the results of Fig.~\ref{fig:acc_f1},
we have pre-processed the data, setting $\mathsf{H}=\mathsf{W}=2^4$, and scaled all pixel values to the range of $-1$ to $1$.
After flattening an image, we obtain a $d=2^8$ component feature vector $\textbf{x}$, which is then transformed into $|\textbf{x}\rangle$ following \eqref{x-kets}. Subsequently, we pass $|\textbf{x}\rangle$ through the INN shown in Fig.~\ref{fig:Int}, utilizing $L=2$ layers.
Other hyperparameters, including the learning rate, betas, batch size, and number of epochs, are set to $0.01$, $(0.5,0.9)$, $2^6$, and $10$, respectively (for additional details, refer to \cite{Arun_GitHub}).
For every $(a,b)$, we maintain the same settings as described above to obtain the results presented in Fig.~\ref{fig:acc_f1}. In every case, the INN achieves an accuracy of more than $80\%$ and $\text{f}_1$ score more than $75\%$. 
These scores can potentially be further improved by modifying the INN architecture and fine-tuning the hyperparameters.

To classify images of the MNIST dataset into the $C=10$ classes, we need a NN that takes a $d_\text{in}$-dimensional input and provides a $d_\text{out}$-dimensional output, where $d_\text{out}$ is not necessarily the same as $d_\text{in}$.
Let us take $d_\text{in}=3$, $d_\text{out}=2$, and compare 
\begin{align}
	\label{W, FU}
	&\textbf{x}'=W\textbf{x} =
	\begin{pmatrix}
		w_{00} & w_{01} & w_{02} \\ 
		w_{10} & w_{11} & w_{12} \\   
	\end{pmatrix} 
	\begin{pmatrix}
		x_0 \\
		x_1 \\
		x_2 
	\end{pmatrix}  
	\quad\text{with}\\
	&FU(\boldsymbol{\phi})|\textbf{x}\rangle \equiv
	\frac{1}{\sqrt{3}}
	\begin{pmatrix}
		\text{e}^{\ii\phi_{0}} & 
		\text{e}^{\ii\phi_{1}} & 
		\text{e}^{\ii\phi_{2}}  
		\\ 
		\text{e}^{\ii\phi_{0}} & 
		\text{e}^{\ii(\phi_{1}+\frac{2\pi}{3})} & 
		\text{e}^{\ii(\phi_{2}+\frac{4\pi}{3})}  
		\\  
		\text{e}^{\ii\phi_{0}} & 
		\text{e}^{\ii(\phi_{1}+\frac{4\pi}{3})} & 
		\text{e}^{\ii(\phi_{2}+\frac{2\pi}{3})}   
	\end{pmatrix} 
	\begin{pmatrix}
		x_0 \\
		x_1 \\
		x_2 
	\end{pmatrix}.  \nonumber
\end{align}
The above equations represent a single linear layer of a classical NN (multi-layer perceptron) \cite{LeCun15,Goodfellow16, Aggarwal23} and a single layer of INN of Fig.~\ref{fig:Int}, respectively.
Here, we have used amplitude encoding \eqref{x-kets} by assuming
$\|\textbf{x}\|=1$. 
The linear layer combines the input features from \textbf{x} in a weighted manner to generate new features in $\textbf{x}'$, while the INN layer creates \emph{interference} among the input features to generate new features.

Two more differences can be observed through \eqref{W, FU}.
Firstly, the output dimension is 2 in the case of the linear layer, while it is $3$ (not equal to the desired $d_\text{out}$) for the INN layer. Secondly, each row of $W$ is \emph{unrelated} to the others, whereas the rows of any unitary matrix 
must follow the \emph{orthogonality relation}.
As a result, we get  
\emph{distinct} new features $\sum_j w_{0j} x_j$ and $\sum_j w_{1j} x_j$ from the linear layer.
Whereas, from the INN layer, the new features like $\sum_j \text{e}^{\ii\phi_{j}} x_j$ and $\sum_j \text{e}^{\ii\phi_{j}} x_j\, \omega^j$ 
represent very similar functions (machine learning modes) of the \emph{same} set $\boldsymbol{\phi}$ of parameters, and $\omega=\text{e}^{\texttt{i}\frac{2\pi}{3}}$ is a constant. Furthermore, their derivatives with respect to a parameter only differ by a constant.
So, to get $d_\text{out}$ \emph{distinct} features, we put $d_\text{out}$ distinct INNs of Fig.~\ref{fig:Int} in \emph{parallel} and obtain
\begin{align}
	\label{pt}
	|\textbf{x}, \Phi^t \rangle & = \textbf{U}(\Phi^t)|\textbf{x}\rangle\,,
	\quad\mbox{where} \nonumber\\
	\textbf{U}(\Phi^t) &= \left(\prod_{l=1}^{L} F U(\boldsymbol{\phi}^{tl}) \right) F
	\quad\mbox{and} \nonumber\\
	p_t	&= |\langle 0|\textbf{x}, \Phi^t \rangle|^2
\end{align}
for $t=0,\cdots,d_\text{out}-1$. 
Similar to how \eqref{p} represents a sequence of interferometers, \eqref{pt} portrays a block of sequences of interferometers, as displayed in Fig.~\ref{fig:Int_blk}.
In the figure, we also illustrate the similarity of \eqref{pt} to a linear layer that represents $d_\text{out}$ linear regression models in \emph{parallel}.
One can observe that, from the same ${|\textbf{x}\rangle}$, we get $d_\text{out}$ distinct probabilities $p_t$ through $d_\text{out}$ \emph{mutually independent} unitary operators $\textbf{U}(\Phi^t)$, each of
which represents a sequence of $L$ interferometers.
Since these probabilities are independent of each other, they are not required to sum up to 1. If their normalization is needed, one can apply the function
\begin{equation}
	\label{softmax}
	\text{softmax}\,(p_t) = \frac{\text{e}^{\,p_t}}%
	{\sum_{t=0}^{d_\text{out}-1}\text{e}^{\,p_t}}
\end{equation}
to every component of $\textbf{p}=(p_0,\cdots,p_{d_\text{out}-1})$.
Essentially, the INN block \eqref{pt}
creates distinct new features $\textbf{p}$
from the old \textbf{x} through independent interferences.
The vector $\textbf{p}$ can serve as a final output or an input to the next INN block as shown in Fig.~\ref{fig:2Blks}.

Figure~\ref{fig:2Blks} shows an INN whose architecture is defined by
$\textbf{d}=(d_0,\cdots, d_{M-1}, d_{M})$ and
$\textbf{L}=(L_0,\cdots, L_{M-1})$, where the number of blocks $M$ is 2.
Its $m$th block takes $d_m$-dimensional input and
gives $d_{m+1}$-dimensional output.
Within the block, there are $d_{m+1}$ parallel sequences of interferometers represented by $\otimes_{t=0}^{d_{m+1}-1} \textbf{U}(\Phi^t)$, and the length of each sequence is $L_m$.

While the INN in Fig.~\ref{fig:2Blks} draws inspiration from multilayer perceptrons (NNs) \cite{LeCun15,Goodfellow16,Aggarwal23} like the quantum NNs (QNNs) in \cite{Farhi18,Altaisky01,Beer20}, it differs in the following aspects.
A single quantum preceptron in \cite{Altaisky01} is a sum of single-qubit unitary operators, whereas it is a product of unitary operators in the case of INN as shown in \eqref{p}.
Compared to \cite{Farhi18,Beer20}, the INN has a parallel structure of unitary operators, as depicted in Fig.~\ref{fig:Int_blk}.
In contrast to the QNNs from \cite{Farhi18,Altaisky01,Beer20}, where each node represents a qubit, INNs do not necessitate a qubit system. INNs are applicable in any dimension and explicitly incorporate Fourier transformations.

\begin{table}
	\centering
	\caption{For the two datasets, we present the accuracy and average $\text{f}_1$ score achieved on their test sets by two separately trained INNs specified by the $\textbf{L}$ values. Both the INNs have, same $\textbf{d}=(2^8, 2^6, C)$, a $M=2$ blocks architecture as depicted in Fig.~\ref{fig:2Blks}. Notably, as we increase the number of layers, both metrics show improvement for both datasets.}
	\label{tab:acc_f1}
\begin{tabular}{c|c|c|c}
	\hline \hline
		dataset     & metric & $\textbf{L}=(1,1)$ & $\textbf{L}=(2,2)$ \\ \hline
	\multirow{2}{*}{MNIST} & accuracy        & 0.89 & 0.93 \\ 
	                 & average $\text{f}_1$  & 0.89 & 0.93 \\ \hline
	\multirow{2}{*}{FashionMNIST} & accuracy & 0.78 & 0.83 \\ 
	                 & average $\text{f}_1$  & 0.77 & 0.83 \\ \hline\hline
\end{tabular}
\end{table}

We have employed INNs, as illustrated in Fig.~\ref{fig:2Blks}, for image classification on both the MNIST and FashionMNIST \cite{Xiao17} datasets, each with $C=10$ classes. The FashionMNIST dataset, like MNIST, consists of images representing clothing items from ten (labeled 0 to 9) different categories.
In both cases, we have adopted $\textbf{d}=(2^8, 2^6, C)$, the amplitude encoding of $16\times16$ images, $0.01$ learning rate, $2^5$ batch size, and 3 epochs (for further details, refer to \cite{Arun_GitHub}).
Their performances on the test sets are presented in Table~\ref{tab:acc_f1} for 
$\textbf{L}=(1,1)$ and $(2,2)$.
As we increase the number of layers, the performance---measured by 
the accuracy and average $\text{f}_1$---of the INN improves for both datasets.
In summary, we have attained accuracies and average $\text{f}_1$ scores of ${93\%}$ and ${83\%}$ on the MNIST and FashionMNIST datasets, respectively.

The $\text{f}_1$ score in \eqref{acc_f1} pertains to the positive class.
We can utilize a similar formula to calculate the $\text{f}_1$ score for each class and then derive the average $\text{f}_1$ score.
It is also possible to extend the accuracy formula presented in \eqref{acc_f1} from $C=2$ to $C=10$ classes.
In the case of the cross-entropy loss function described in \eqref{loss}, it is essential to ensure that the probabilities are appropriately normalized. This can be accomplished by employing the softmax function outlined in \eqref{softmax}.

We conclude this section with the following two observations:
(i) It is of interest to investigate whether the pair $\{U(\boldsymbol{\phi}), F\}$ from \eqref{FU} constitutes a universal set---capable of generating any unitary operation through multiplications---for quantum computation with a $d$-level system. Notably, for $d=2$, it is a known universal set \cite{Nielsen_Chuang}. Moreover, $U(\boldsymbol{\phi})$ encompasses all diagonal unitary operations and can be expressed as a linear combination of different powers of $Z=\sum_{j=0}^{d-1} \omega^j |j \rangle\langle j|$. 
Through multiplication, $Z$ and $X=F^3 Z F$ can generate the 
Heisenberg-Weyl group \cite{Durt10}, whose elements constitute the unitary operator bases \cite{Schwinger60}.

(ii) As a black and white image is a two-dimensional (2D) object, instead of \eqref{x-kets}, one can use
\begin{equation}
	\label{x-kets-2}
	|\textbf{x}\rangle_\text{2D}  = 
	\frac{1}{\|\textbf{x}\|}\sum_{j=0}^{\mathsf{H}-1}  \sum_{j'=0}^{\mathsf{W}-1} x_{jj'}\,
	|j \rangle\otimes|j' \rangle\,,
\end{equation}
to encode an image \textbf{x} into two quantum subsystems with dimensions $\mathsf{H}$ and $\mathsf{W}$, where $d=1\times\mathsf{H}\times\mathsf{W}$. Then, one can use the tensor product $F_\mathsf{H}\otimes F_\mathsf{W}$ and
$U(\boldsymbol{\phi}) = \sum_{k}\sum_{k'}\text{e}^{\texttt{i}\phi_{kk'}}
|k \rangle\langle k|\otimes|k' \rangle\langle k'|$
at the place of \eqref{FU}. 
The tensor product performs a {2D} discrete Fourier transform on the matrix $x_{jj'}$,
with $F_\mathsf{H}$ and $F_\mathsf{W}$ operating exclusively on their respective subsystems.
Afterward, $U(\boldsymbol{\phi})$ executes a componentwise multiplication between the phase matrix (filter) $\text{e}^{\texttt{i}\phi_{kk'}}$ and the Fourier-transformed matrix $\widehat{x}_{kk'}$.
As per the convolution theorem, such multiplication in the frequency domain is linked to \emph{convolution} in the spatial domain through Fourier transforms.
Fourier transforms are fundamental components of digital image processing, particularly for designing various filters in the frequency domain \cite{Gonzalez02}.

%===========================================
\section{IGAN}\label{sec:IGAN}

\begin{figure}
	\includegraphics[width=0.45\textwidth]{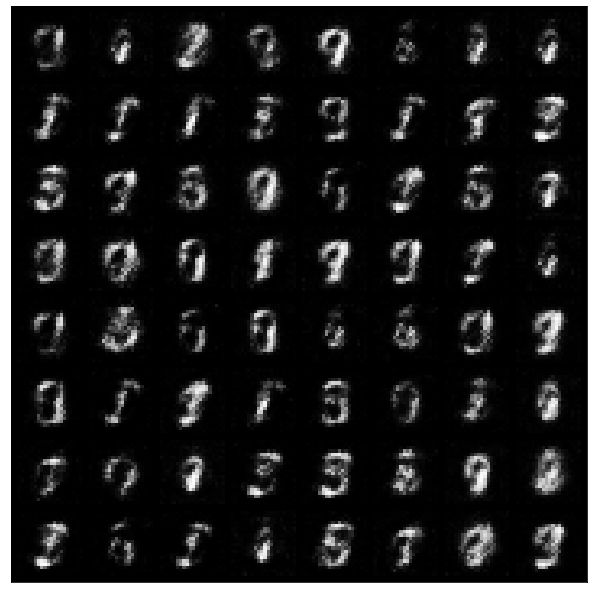}\vspace{2mm}
	\includegraphics[width=0.45\textwidth]{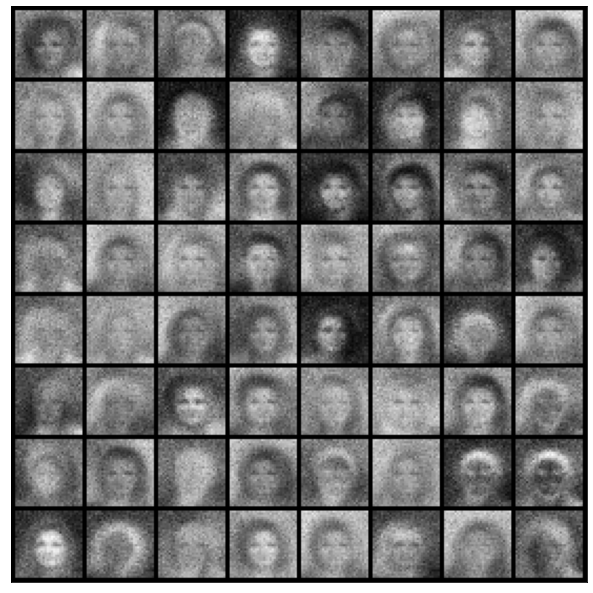}
	\caption{After separately training IGANs on the MNIST \cite{MNIST} and CelebA \cite{CelebA} datasets, the top and bottom panels showcase {64} \emph{fake} images generated by their respective generators. To enhance contrast, we have employed the transformation $\textbf{p}\rightarrow \tanh(3(2\,\textbf{p}-1))$ at the place of \eqref{tanh} exclusively for image generation in the top panel and Fig.~\ref{fig:0-9}.
	}
	\label{fig:gen_images}
\end{figure}

\begin{figure}
	\includegraphics[width=0.45\textwidth]{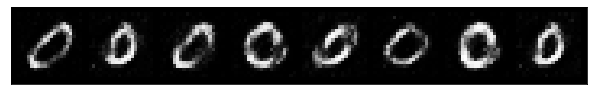}\\\vspace{-2mm}
	\includegraphics[width=0.45\textwidth]{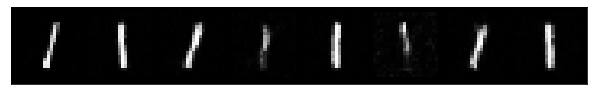}\\\vspace{-2mm}
	\includegraphics[width=0.45\textwidth]{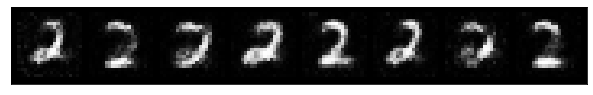}\\\vspace{-2mm}
	\includegraphics[width=0.45\textwidth]{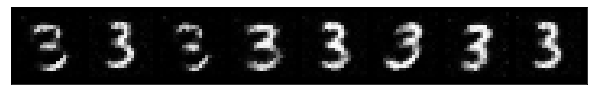}\\\vspace{-2mm}
	\includegraphics[width=0.45\textwidth]{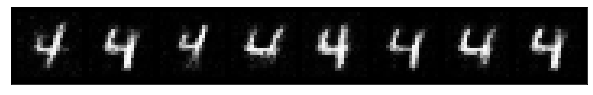}\\\vspace{-2mm}
	\includegraphics[width=0.45\textwidth]{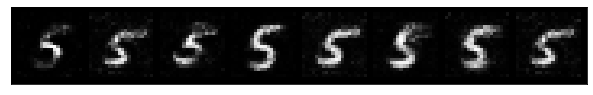}\\\vspace{-2mm}
	\includegraphics[width=0.45\textwidth]{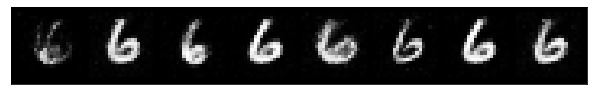}\\\vspace{-2mm}
	\includegraphics[width=0.45\textwidth]{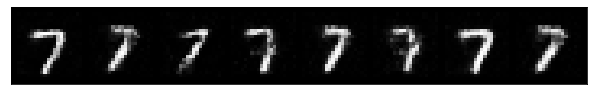}\\\vspace{-2mm}
	\includegraphics[width=0.45\textwidth]{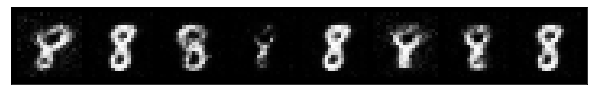}\\\vspace{-2mm}
	\includegraphics[width=0.45\textwidth]{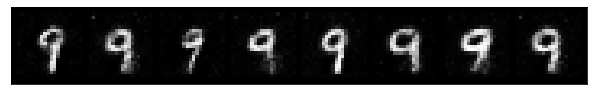}
	\caption{From top to bottom, each row displays eight \emph{fake} images generated by the $G$ network trained separately on real images of digits $0,\cdots,9$. For each digit, we maintained consistent architecture and hyperparameters (with the exception of data size) as those presented in Tables \ref{tab:arch} and \ref{tab:hyper} for the MNIST dataset. The training data was sourced from the MNIST training set.}
	\label{fig:0-9}
\end{figure}

\begin{figure}
	\includegraphics[width=0.35\textwidth]{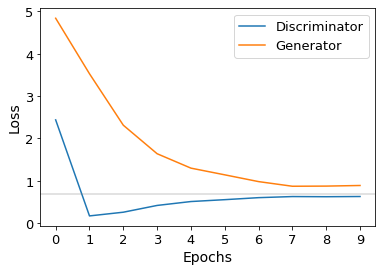}\\
	\includegraphics[width=0.35\textwidth]{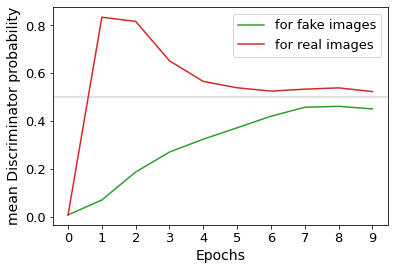}
	\caption{At the top, we present the evolution of $\mathcal{L}_D$ and $\mathcal{L}_G$ at various epochs, represented by orange and blue curves, respectively. Meanwhile, at the bottom, we illustrate the changes in $\mathcal{P}_\text{real}$ and $\mathcal{P}_\text{fake}$ over different epochs, denoted by red and green graphs, respectively. The definitions of these loss functions and average probabilities can be found in \eqref{loss_GAN} and \eqref{mean_probs}, respectively. 
	In the top and bottom plots, the gray horizontal lines denote the values 
	$\ln(2)$ and $\frac{1}{2}$, respectively.
	Both the plots pertain to the IGAN trained on the CelebA dataset.
}
	\label{fig:loss_prob}
\end{figure}

In this section, we introduce IGANs, and our primary contributions are summarized as follows: Tables~\ref{tab:arch} and \ref{tab:hyper} detail the architectures and hyperparameters of IGANs, Algorithm~\ref{Algo:IGAN} outlines the training process, and the implementation can be found in \cite{Arun_GitHub}. 
Figure~\ref{fig:loss_prob} offers insights into how losses and probabilities evolve during the training. And, Figs.~\ref{fig:gen_images} and \ref{fig:0-9} display sample images generated after the training process.

In 2014, Ian Goodfellow and colleagues introduced the generative adversarial networks (GANs) \cite{Goodfellow14}, comprising two NNs: the generator $G$ and the discriminator $D$.
The generator takes random noise vectors $\textbf{z}$ and endeavors to produce outputs $G(\textbf{z})$ that closely resemble real data. 
Meanwhile, the discriminator, acting as a binary classifier, distinguishes between \emph{real} data ${\textbf{x}}$ labeled as 1 and \emph{fake} data ${G(\textbf{z})}$ labeled as 0. It assesses both real and generated data and provides a probability score for an input being real.

To enhance classification accuracy, the discriminator aims to drive $D(\textbf{x})$ closer to 1 through maximization and $D(G(\textbf{z}))$ closer to 0 through minimization. Both objectives are accomplished by minimizing the discriminator's loss
\begin{align}
	\label{loss_GAN}
	\mathcal{L}_D& = -\frac{1}{2N}\sum_{i=1}^N\left[\ln(D(\textbf{x}^{[i]})) + \ln(1-D(G(\textbf{z}^{[i]})))\right]\,,
	 \nonumber\\
    \mathcal{L}_G& = -\frac{1}{N}\sum_{i=1}^N\ln(D(G(\textbf{z}^{[i]})))
\end{align}
serves as a guide for the generator. It encourages the generator to produce increasingly realistic data, thereby pushing $D(G(\textbf{z}))$ closer to 1.
In \eqref{loss_GAN}, every summation  is over a mini-batch,
where $\{\textbf{x}^{[i]}\}_{i=1}^N$ originates from the given data and $\{\textbf{z}^{[i]}\}_{i=1}^N$ is drawn from a prior (in our case, the standard normal) distribution.

The two networks are trained simultaneously as described in Algorithm~\ref{Algo:IGAN} inspired from \cite{Goodfellow14}. Initially, $\mathcal{L}_D$ is minimized with respect to the discriminator's parameters, followed by minimizing $\mathcal{L}_G$ with respect to the generator's parameters within a single iteration (epoch). 
The two networks compete with each other and seek the Nash equilibrium point, where the discriminator is unable to distinguish between real and fake data. There both the probabilities $D(\textbf{x})$ and $D(G(\textbf{z}))$ attain a value of $\frac{1}{2}$, resulting in $\mathcal{L}_D=\ln(2)=\mathcal{L}_G$.

Training a GAN can be a challenging and unstable process. Therefore, several GAN variants, such as Deep Convolutional GANs \cite{Radford15} and Wasserstein GANs \cite{Arjovsky17,Gulrajani17}, have been developed. Deep Convolutional GANs introduced architectural guidelines for generator and discriminator networks, leading to more stable training and the generation of realistic and detailed images. Wasserstein GANs introduced the Wasserstein distance as a more stable and meaningful loss function for GAN training, effectively addressing issues like mode collapse.

\begin{algorithm}%[H]
	\caption{IGAN training algorithm}
	\label{Algo:IGAN}
	\textbf{Input}: batch size $N$, number of epochs, learning rate,
	Adam hyperparameters $(\beta_1, \beta_2)$ from Table~\ref{tab:hyper}.
	
	$\bullet$ Initialize parameters of discriminator $\boldsymbol{\Phi}_D$ and of generator $\boldsymbol{\Phi}_G$ with the Xavier initialization \cite{Glorot10}.
	
	\For{number of epochs}{
		\For{number of mini-batches}{
			$\bullet$ Sample mini-batch $\{\textbf{z}^{[i]}\}_{i=1}^N$ from the
			standard normal distribution
			${\mathcal{N}(0,1)}$.
			
			$\bullet$ Sample mini-batch $\{\textbf{x}^{[i]}\}_{i=1}^N$ from the training dataset.
			
			$\bullet$ Update 
			$\boldsymbol{\Phi}_D \leftarrow 
			\text{Adam}(\nabla_{\boldsymbol{\Phi}_D} \mathcal{L}_D, \beta_1, \beta_2)$
			
			$\bullet$ Update 
			$\boldsymbol{\Phi}_G \leftarrow 
			\text{Adam}(\nabla_{\boldsymbol{\Phi}_G} \mathcal{L}_G, \beta_1, \beta_2)$
		}
	}
\end{algorithm}

\begin{table}
	\centering
	\caption{The table presents the architecture of INNs for the discriminator $D$  and generator $G$ of separate IGANs trained on the two datasets. In both cases, every INN is of the structure depicted in Fig.~\ref{fig:2Blks}.}
	\label{tab:arch}
	\begin{tabular}{c|c|c|c}
		\hline \hline
		dataset     & INN & $\textbf{d}$ & $\textbf{L}$ \\ \hline
		\multirow{2}{*}{MNIST}  & $D$    & $(20^{2}, 2^6,  1)$   & $(3,3)$ \\ 
		& $G$    & $(2^3, 2^6,20^{2})$   & $(3,3)$ \\ \hline
		\multirow{2}{*}{CelebA} & $D$    & $(32^{2}, 2^7,  1)$   & $(3,3)$ \\
		& $G$    & $(2^5, 2^7,  32^{2})$ & $(3,3)$ \\ \hline\hline
	\end{tabular}
\end{table}

\begin{table}
	\centering
	\caption{The table presents the hyperparameters utilized in Algorithm~\ref{Algo:IGAN} for training separate IGANs on the two datasets.}
	\label{tab:hyper}
	\begin{tabular}{c|c|c|c|c|c}
		\hline \hline
		dataset     & data size & batch size & epochs & rate & $(\beta_1,\beta_2)$ \\ \hline
		MNIST       & $5000$   & $2^7$  & 10 & $0.01$ & $(0.5,0.9)$ \\  \hline
		CelebA      & $50000$   & $2^9$ & 10 & $0.01$ & $(0.5,0.9)$ \\ \hline\hline
	\end{tabular}
\end{table}

In the quantum domain, several models have been employed for image generation, including the quantum GANs \cite{Huang21, Tsang22, Zhou23, Stein21, Chu22},
Born machines \cite{Liu18, Benedetti19b, Rudolph22},
matrix product states \cite{Han18}, and
quantum variational autoencoder \cite{Khoshaman19}.
In quantum GANs, described in \cite{Huang21, Tsang22}, images are created from multiple patches generated in parallel by sub-generators from a product state, which encodes the components of \textbf{z} into rotation angles.
The loss function in \cite{Tsang22} is inspired by Wasserstein GANs \cite{Gulrajani17}. In contrast, quantum GANs in \cite{Stein21, Chu22} leverage quantum state fidelity-based loss functions and incorporate principal component analysis for input image compression.
In the case of hybrid quantum-classical GANs discussed in \cite{Zhou23}, a specific remapping method is employed to enhance the quality of generated images.
The generative models in \cite{Liu18, Benedetti19b, Han18} are built using datasets of binary images. 
In \cite{Rudolph22, Benedetti18, Khoshaman19}, quantum models are employed for producing noise vectors \textbf{z}.
Similar to our IGAN, the MNIST dataset is utilized in all the generative models introduced in \cite{Huang21, Tsang22, Zhou23, Stein21, Chu22, Han18, Rudolph22, Benedetti18, Khoshaman19}.

Now, we introduce our IGANs, where both the generator $G$ and discriminator $D$ are INNs as shown in Fig.~\ref{fig:2Blks}.
It is important to note that our approach does not involve principal component analysis for input compression, binary images for training, or the inclusion of any classical layers within our INNs.
For input loading, we always use amplitude encoding \eqref{x-kets} for classical vectors such as \textbf{x}, \textbf{z}, and \textbf{p}.
An output from an interferometric block is always a classical vector \textbf{p} as illustrated in Fig.~\ref{fig:Int_blk}. 
From the generator, we get $\textbf{p}\in[0,1]^{\times d}$ with an image size of  $d=\mathsf{C}\times\mathsf{H}\times\mathsf{W}$, which,
after the componentwise transformation 
\begin{equation}
	\label{tanh}
\textbf{p}\rightarrow \tanh(2\,\textbf{p}-1)=G(\textbf{z})\in[-1,1]^{\times d}\,,
\end{equation}
is fed to the discriminator during training. 
The discriminator's output, $D(\textbf{x})$ or $D(G(\textbf{z}))$, represents the probability, $p\in[0,1]$, of the input being real, where the corresponding $\textbf{p}=(1-p\,,p)$. The training process for our IGANs is outlined in Algorithm~\ref{Algo:IGAN} and its complete implementation is given in \cite{Arun_GitHub}.

\begin{figure}
	\includegraphics[width=0.45\textwidth]{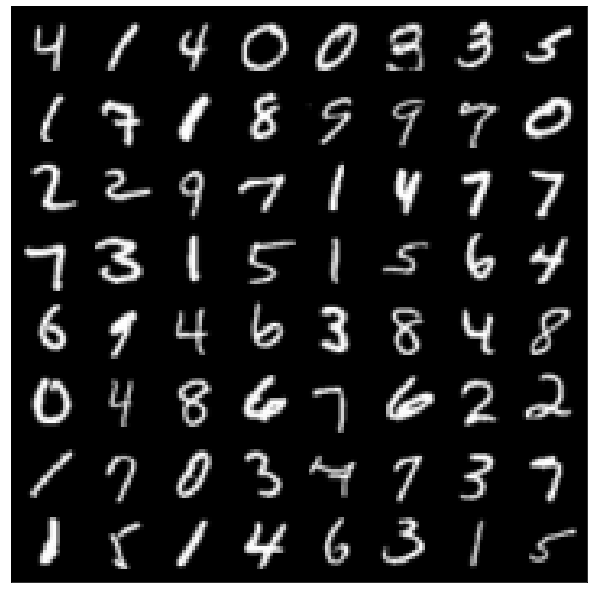}\vspace{2mm}
	\includegraphics[width=0.45\textwidth]{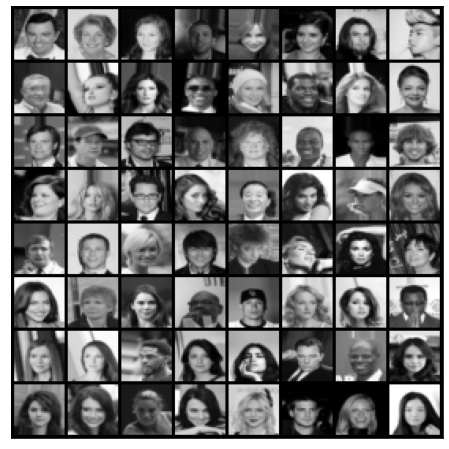}
	\caption{The top and bottom panels showcase samples of {64} \emph{real} images from the MNIST \cite{MNIST} and CelebA \cite{CelebA} datasets, respectively.}
	\label{fig:real_images}
\end{figure}

Separate IGANs were trained on the MNIST dataset \cite{MNIST} and CelebA datasets \cite{CelebA}, which consists of images featuring celebrities' faces
as displayed in Fig.~\ref{fig:real_images}.
Subsequently, samples of generated images from their respective $G$s are presented in Fig.~\ref{fig:gen_images}. In both cases, black and white images were used for training, and 2-block INNs were employed for both $D$ and $G$.
For the two datasets, we have adopted slightly different architectures for the INNs, as detailed in Table~\ref{tab:arch}, and slightly varied hyperparameters, as specified in Table~\ref{tab:hyper}.

In the case of the MNIST dataset, we have randomly taken {5000} images of 0 to 9 digits for the training as specified in the ``data size'' column of Table~\ref{tab:hyper}. Each image has dimensions  $d=\mathsf{C}\times\mathsf{H}\times\mathsf{W}=1\times20\times20$, as evident from the first and last components of \textbf{d}s for $D$ and $G$ in Table~\ref{tab:arch}.
In the top panel of generated images in Fig.~\ref{fig:gen_images}, clear representations of numbers 0, 1, 3, 5, and 9 are easily noticeable. However, the images of 2, 4, and 6 appear less distinct, while those of 7 and 8 do not seem to be present.
In another set of experiments, we trained the same IGAN using images of a single digit at a time and achieved the results depicted in Fig.~\ref{fig:0-9}.
These results demonstrate our capability to generate images corresponding to all ten digits, from 0 to 9. One can visually assess the quality of generated images by comparing them to real images through Figs.~\ref{fig:real_images}--\ref{fig:0-9}.

In the case of the CelebA dataset, we randomly selected {50000} images for training. These images were converted from color to black and white and resized to the dimensions of $d=1\times32\times32$, as detailed in Table~\ref{tab:arch}. With the table, one can also compute that the INNs for $D$ and $G$ have a total of {393600} and {405504} trainable parameters (phases), respectively.
In Fig.~\ref{fig:loss_prob}, we depict how the discriminator and generator losses converge toward $\ln(2)$ as training progresses. Additionally, we illustrate how the average discriminator probabilities
\begin{equation}
	\label{mean_probs}
	\mathcal{P}_\text{real}=\frac{1}{N}\sum_{i=1}^N D(\textbf{x}^{[i]})
	\quad \mbox{and}\quad
	\mathcal{P}_\text{fake}=\frac{1}{N}\sum_{i=1}^N D(G(\textbf{z}^{[i]}))
\end{equation}
approach $\tfrac{1}{2}$. After the training, we got the ability to generate 
images featuring human faces, as showcased in Fig.~\eqref{fig:gen_images}.
By the way, with the average probabilities, one can define
$\mathcal{L}'_D=-\mathcal{P}_\text{real}+\mathcal{P}_\text{fake}$ and $\mathcal{L}'_G=-\mathcal{P}_\text{fake}$, inspired from the Wasserstein loss functions \cite{Arjovsky17, Gulrajani17, Tsang22}, at place of \eqref{loss_GAN}.

%===========================================
\section{conclusion and outlook}\label{sec:conclusion}

We have introduced INNs, which are artificial neural networks composed of interferometers.
An INN is a sequence of interferometric blocks, with each block containing parallel sequences of interferometers. 
While one could, in principle, use an INN with classical waves, it can be regarded as a quantum model, devoid of any classical layers.

We have shown that INNs are useful for optimization as well as supervised or unsupervised machine learning tasks. 
For the QUBO problems, we achieve the global minimum approximately ${90\%}$ of the time, with the remaining instances typically falling within a range of ${2\%}$ from the global minimum.
In the context of multi-class image classification problems, we achieved accuracies of
${93\%}$ and ${83\%}$ on the MNIST and FashionMNIST datasets, respectively.
While our accuracy falls short in comparison to state-of-the-art classical NNs, which provide ${99.87\%}$ accuracy on the MNIST dataset \cite{Byerly20}, and ${96.91\%}$ accuracy on the FashionMNIST dataset \cite{Tanveer20}, it is important to note that INNs exhibit a simpler architecture. Nonetheless, our work represents a significant step forward in the development of more advanced quantum NNs.

We have introduced IGANs, made of INNs, for image generation. These IGANs have successfully generated images of 0 to 9 digits and human faces.
While our image quality may currently lag behind that of classical GANs \cite{Goodfellow14, Radford15, Arjovsky17, Gulrajani17}, there is potential for enhancement through network modifications, architectural adjustments, and fine-tuning of hyperparameters. 
Last but not least, it is crucial to analyze the robustness of our INNs against noise and qubit errors, which will be the focus of future research.

%\vfill
%\bigskip
%\clearpage

%===========================================

%===========================================

\end{document}